\documentclass[a4paper,11pt]{article}
\usepackage{jheppub} 
\usepackage{lineno}
\usepackage{booktabs}

\usepackage{amsmath}
\usepackage{amssymb}
\usepackage{braket}
\usepackage{dsfont}
\usepackage{bm}
\usepackage{physics}
\usepackage{stmaryrd} 
\usepackage{cancel} 
\usepackage{siunitx}
\usepackage{float}
\usepackage{graphicx}
\usepackage{subcaption}
\usepackage{mathtools}
\usepackage{algorithm}
\usepackage{algpseudocode}
\usepackage{booktabs}
\usepackage{xcolor}
\usepackage[vvarbb, subscriptcorrection]{newtx}
\usepackage{tikzit}

\tikzstyle{new style 0}=[fill={rgb,255: red,128; green,128; blue,128}, draw=black, shape=circle]
\tikzstyle{new style 1}=[fill={rgb,255: red,0; green,176; blue,0}, draw=black, shape=circle]
\tikzstyle{new style 2}=[fill=blue, draw=black, shape=circle]
\tikzstyle{new style 3}=[fill=white, draw=black, shape=circle]

\tikzstyle{new edge style 0}=[draw=black, ->, thick]
\tikzstyle{new edge style 1}=[->, draw=red, thick]
\tikzstyle{new edge style 2}=[-, thick]
\tikzstyle{new edge style 3}=[-, draw=red, thick]
\tikzstyle{new edge style 4}=[-, dashed, thick, draw=red]
\tikzstyle{new edge style 5}=[thick, dashed, ->]
\tikzstyle{new edge style 6}=[->, thick, dashed, draw=red]
\tikzstyle{new edge style 7}=[-, draw=blue, thick]
\tikzstyle{new edge style 8}=[draw=blue, ->, thick, dashed]
\tikzstyle{new edge style 9}=[draw={rgb,255: red,128; green,128; blue,128}, thick, dashed, ->]
\tikzstyle{new edge style 10}=[draw={rgb,255: red,0; green,176; blue,0}, thick, ->]
\tikzstyle{new edge style 11}=[draw=blue, thick, ->]
\tikzstyle{new edge style 12}=[-, thick, draw={rgb,255: red,0; green,176; blue,0}]

\usepackage[bf]{titlesec}
\titlelabel{\thetitle.\quad}
\newcommand{\ee}{\mathrm{e}}
\newcommand{\iu}{\mathrm{i}}
\newcommand{\pll}[1]{[#1]^+}
\newcommand{\pmm}[1]{[#1]^-}

\renewcommand{\Tr}[1]{\mathrm{Tr}\left[ #1 \right]}

\renewcommand{\Re}{\operatorname{Re}}
\renewcommand{\Im}{\operatorname{Im}}

\newcommand{\pfact}{p^{\mathrm{fact}}}
\newcommand{\thetar}{\theta_\mathrm{refresh}}
\newcommand{\thetas}{\theta_\mathrm{sample}}
\newcommand{\thetae}{\theta_\mathrm{event}}
\newcommand{\Thetar}{\Theta_\mathrm{refresh}}
\newcommand{\Thetas}{\Theta_\mathrm{sample}}

\newcommand{\shat}[1]{\smash{\hat{#1}}}

\title{Event-Chain Monte Carlo for Yang-Mills SU(N) lattice field theory I : Design and proof of concept}

\author[a,1]{Benoît Blossier,\note{blossier@ijclab.in2p3.fr}}
\author[b,2]{Manon Michel,\note{manon.michel@uca.fr}}
\author[a,3]{Yacob Ozdalkiran\note{ozdalkiran@ijclab.in2p3.fr}}
\affiliation[a]{Laboratoire de Physique des 2 Infinis Irène Joliot-Curie, CNRS/IN2P3, \\Université Paris-Saclay, 91405 Orsay Cedex, France}
\affiliation[b]{Laboratoire de Mathématiques Blaise Pascal UMR 6620, CNRS, \\Université Clermont-Auvergne, Aubière, France}

\abstract{We develop two implementations of the Event-Chain Monte Carlo (ECMC) algorithm for Yang-Mills $\mathrm{SU}(N)$ lattice gauge theories with the Wilson action.
	These algorithms consist in a succession of local ballistic updates intersped with stochastic events, resulting in an irreversible and rejection-free Markov process.
	The resulting dynamics satisfy global balance, ensuring the correct equilibrium distribution. The algorithms are formulated for general $\mathrm{SU}(N)$ Yang-Mills theories with Wilson action and implemented for the case $N=3$. Numerical tests on four-dimensional lattices show that standard gauge observables, such as the mean plaquette, agree with results obtained using conventional Monte Carlo algorithms. These results provide a first validation of ECMC as a viable sampling scheme for Yang-Mills lattice gauge theories.}

\makeatletter
\gdef\@fpheader{}
\makeatother
\begin{document}
	\maketitle
	\flushbottom
	\newpage
	\section{Introduction}
	Lattice gauge theory provides the only systematically improvable framework for the non-perturbative study of Quantum Chromodynamics (QCD). In this approach, expectation values of observables are computed by sampling of gauge field configurations distributed according to the Boltzmann factor appearing in the Euclidean path integral. In the case of pure gauge theories described by the Wilson action \cite{wilsonConfinementQuarks1974} and its $O(a^2)$ improved formulations, a variety of Markov Chain Monte Carlo (MCMC) algorithms have been developed over the past decades, including local update schemes such as the Heatbath \cite{kennedyImprovedHeatbathMethod1985} and over-relaxation \cite{creutzOverrelaxationMonteCarlo1987} algorithms, as well as global approaches such as Hybrid Monte Carlo (HMC) \cite{duaneHybridMonteCarlo1987} when fermions are present. While these methods have proven highly successful, they remain affected by critical slowing down as the lattice spacing is finer and finer \cite{schaeferCriticalSlowingError2011, allesHybridMonteCarlo1996}. In particular, autocorrelation time of the topological charge diverges in the continuum limit, meaning that topological modes become increasingly difficult to sample efficiently.
	
	Motivated by these limitations, a considerable effort has been devoted to exploring alternative sampling strategies that modify the dynamical properties of the underlying Markov process. Among them, Open Boundary Conditions (OBC) \cite{luscherLatticeQCDOpen2013} and Parallel Tempering on Boundary Conditions (PTBC) \cite{jooParallelTemperingLattice1999}, biasing and density-of-states techniques like PT-MetaD \cite{eichhornParallelTemperedMetadynamics2025} and Log-Likelyhood Ratio (LLR) \cite{cossuErgodicityLLRMethod2018}, as well as Master-Field simulations \cite{francisMasterfieldSimulationsOaimproved2020} and generative approaches via Normalizing Flows \cite{abbottNormalizingFlowsLattice2023}. While promising, these methods remain active areas of research and development to fully overcome topological freezing.
	
	Another direction explored in this paper consists in relaxing the requirement of detailed balance while preserving global balance, the only necessary condition to ensure the invariance of the target probability distribution. Algorithms in this class generate irreversible Markov chains exhibiting dynamical persistency in configuration space, which can lead to significantly improved sampling efficiency \cite{diaconisAnalysisNonreversibleMarkov2000}. Among such methods, Event-Chain Monte Carlo (ECMC) \cite{michelGeneralizedEventchainMonte2014}	has received increasing attention in recent years. Originally introduced for systems of classical particles, 
	ECMC is based on a lifted Markov process in which the dynamics proceeds through continuous updates interrupted by stochastic “events”. The resulting algorithm satisfies global balance but violates detailed balance. Thus it allows for persistent directed motion in configuration space. ECMC has been successfully applied to a variety of statistical physics models, including hard-sphere systems \cite{bernardEventchainMonteCarlo2009}, lattice spin models \cite{michelEventchainMonteCarlo2015}, bosonized one-dimensional quantum systems \cite{bouverot-dupuisBosonizedOnedimensionalQuantum2025} and asymptotically free lattice field theories \cite{hasenbuschTestingEventchainAlgorithm2018} where substantial reductions in autocorrelation times have been observed.
	
	The extension of Event-Chain Monte Carlo algorithm to lattice gauge theories presents several conceptual and practical challenges. Gauge fields take values in compact non-Abelian Lie groups and the lattice action couples link variables through closed paths. It leads to a highly constrained configuration space. Designing continuous updates interspersed with stochastic events that respect the group structure while maintaining the correct equilibrium distribution is therefore non-trivial. Although Event-Chain algorithms have been explored in several lattice field theory contexts \cite{hasenbuschTestingEventchainAlgorithm2018}, a formulation of ECMC suitable for Yang-Mills gauge theories remains largely unexplored.
	
	In this paper we present an adaptation of Event-Chain Monte Carlo for Yang-Mills SU($N$) lattice gauge theories regularized by the Wilson action. The construction relies on the factorization of the gauge action into plaquette contributions and on continuous updates of link variables along a one-parameter subgroups of the gauge group permitted by the XY-Embedding \cite{manaMultiGridMonteCarlo1997}.
	
	We develop two variations of the Event-Chain algorithm for SU($N$) lattice gauge theories and discuss their implementation in the case of SU(3), which is the gauge lattice QCD in quenched approximation. The first variation, denoted Reflective ECMC, can easily be generalized to any lattice field theory with factorizable action but allows backtracking. We built on the Forward generalization \cite{michelForwardEventChainMonte2020} to propose a second variant that exploits symmetries of the system itself to avoid any backtracking. More precisely, this second Forward version uses the structure of the Wilson action and in particular the symmetry properties of closed loops, thus extending naturally to $O(a^2)$ improved actions.\\
	
	The paper is organized as follows. Section \ref{sec:wilson} presents the formalism of Yang-Mills SU($N$) lattice gauge theory for completeness and introduces the Wilson action with the relevant objects. Section \ref{sec:ecmc} presents the Event-Chain algorithm using a discrete infinitesimal approximation for clarity, followed by the details of our two implementations, namely Reflective ECMC in subsection \ref{sec:rev} and Forward ECMC in subsection \ref{sec:irrev}. Section \ref{sec:pdmp} provides the formulation of our algorithms in the rigorous Piecewise Deterministic Markov Process (PDMP) formalism and the proof of convergence. Finally, section \ref{sec:results} reports first numerical tests for quenched lattice QCD defined by the Wilson action, it presents the mean value and autocorrelation time of the plaquette (a short-distance observable) and compares results to existing algorithms to ensure correctness of our implementations.
	We conclude in Section \ref{Conclusion}. Appendix \ref{sec:details} provides details of the runs presented in Section \ref{sec:results}.
	
	\section{Wilson lattice gauge theory}\label{sec:wilson}

	We consider a pure $\mathrm{SU}(N)$ lattice gauge theory defined on an Euclidean time four-dimensional hyper cubic lattice $\Lambda$ with periodic boundary conditions. The fundamental degrees of freedom are group-valued link variables
	\begin{equation}
		U_{x,\mu} \in \mathrm{SU}(N),
	\end{equation}
	associated with oriented links connecting a lattice site $x$ to a neighboring site $x+\hat{\mu}$, where $\mu = 1,\ldots,4$ denotes the lattice directions and $\hat\mu$ the associated canonical base vector. From now on we assume the lattice spacing $a=1$ for clear notations. Gauge transformations act locally at lattice sites according to
	\begin{equation}
		U_{x,\mu} \;\rightarrow\; \Omega_x \, U_{x,\mu} \, \Omega_{x+\hat{\mu}}^\dagger,
		\qquad
		\Omega_x \in \mathrm{SU}(N).
	\end{equation}
	By definition, any gauge-invariant observable remains unchanged under that transformation.
	
	The elementary gauge-invariant objects are the plaquettes, defined as traces of the ordered product of four link variables around a square of length 1:
	\begin{equation}
		U_{x,\mu\nu}
		=
		U_{x,\mu}
		U_{x+\hat{\mu},\nu}
		U_{x+\hat{\nu},\mu}^{\dagger}
		U_{x,\nu}^{\dagger}.
	\end{equation}
	In the Wilson regularization, the dynamics of the theory is governed by the action
	\begin{equation}
		S[U]=\beta\sum_{x \in \Lambda}\sum_{\mu<\nu}\left(1 - \frac{1}{N}\mathrm{Re}\,\Tr{U_{x,\mu\nu}}\right),
	\end{equation}
	where $\beta = 2N/\varg^2$ is the inverse bare coupling. The partition function is given by
	\begin{equation}
		Z=\int \mathcal{D} U\, \ee^{-S[U]},
	\end{equation}
	where $\mathcal{D}U = \prod_{x,\mu}\dd U_{x,\mu}$, the product of Haar measures on $\mathrm{SU}(N)$.
	
	Expectation values of observables $O[U]$ are computed as
	\begin{equation}
		\langle O \rangle =\frac{1}{Z} \int \mathcal{D} U\,O[U]\, \ee^{-S[U]}.
	\end{equation}
	In lattice simulations these expectation values are estimated through importance sampling of gauge configurations distributed according to $\pi$ such that
	\begin{align} \label{eq:pi}
		\dd\pi(U) = \frac{1}{Z} \ee^{-S[U]}\mathcal{D} U.
	\end{align}
	Direct sampling of independent gauge configurations distributed according to \eqref{eq:pi} cannot be done because of the huge dimensions of the configuration space and the complex form of the action. To sample gauge configurations efficiently, one uses Markov Chain algorithms designed to iteratively evolve an initial state towards the equilibrium distribution by performing either local updates such as Heatbath \cite{kennedyImprovedHeatbathMethod1985} or Metropolis \cite{creutzMonteCarloStudy1980}, which update individual links, or perform global moves like Hybrid Monte Carlo (HMC) \cite{duaneHybridMonteCarlo1987} which propose transitions for the entire lattice simultaneously. The Event-Chain algorithm is a local algorithm, however able to produce collective updates thanks to the persistency of the generated dynamics.
	
	For algorithms that update individual link variables, it is convenient to express the action in terms of the local contribution associated with a given link $U_{x,\mu}$. When updating a link $U_{x,\mu} \rightarrow U_{x,\mu}'$, the difference in action depends only on the links of the surrounding plaquettes
	\begin{equation}\label{eq:deltas}
		\Delta S(U_{x,\mu}\rightarrow U_{x,\mu}') = -\frac{\beta}{N} \Re\Tr{(U_{x,\mu}' - U_{x,\mu})V_{x,\mu}},
	\end{equation}
	where the local term depends on the neighboring links through the so-called staple matrix
	\begin{equation}\label{eq:staples}
		V_{x,\mu}
		=
		\sum_{\nu\neq\mu}
		\left(
		U_{x+\shat{\mu},\nu}
		U_{x+\hat{\nu},\mu}^{\dagger}
		U_{x^{ },\nu}^{\dagger}
		+
		U_{x+\shat{\mu}-\shat{\nu},\nu}^{\dagger}
		U_{x-\shat{\nu},\mu}^{\dagger}
		U_{x-\shat{\nu},\nu}
		\right) = \sum_{j=1}^{2(d-1)}V^j_{x^{ },\mu}.
	\end{equation}
	
\begin{figure}[htbp]
		\centering
		\scalebox{1.5}{\begin{tikzpicture}
	\begin{pgfonlayer}{nodelayer}
		\node [style=none] (0) at (0, 0) {};
		\node [style=none] (1) at (1, 0) {};
		\node [style=none] (2) at (2, 0) {};
		\node [style=none] (3) at (2, 1) {};
		\node [style=none] (4) at (2, 2) {};
		\node [style=none] (5) at (1, 2) {};
		\node [style=none] (6) at (0, 2) {};
		\node [style=none] (7) at (0, 1) {};
		\node [style=none] (8) at (0, -1) {};
		\node [style=none] (9) at (0, -2) {};
		\node [style=none] (10) at (1, -2) {};
		\node [style=none] (11) at (2, -2) {};
		\node [style=none] (12) at (2, -1) {};
		\node [style=none] (13) at (-1, -0.25) {};
		\node [style=none] (14) at (-2, -0.5) {};
		\node [style=none] (15) at (1, -0.25) {};
		\node [style=none] (16) at (0, -0.5) {};
		\node [style=none] (17) at (3, 0.25) {};
		\node [style=none] (18) at (4, 0.5) {};
		\node [style=none] (19) at (1, 0.25) {};
		\node [style=none] (20) at (2, 0.5) {};
		\node [style=none] (21) at (3, 0.5) {};
		\node [style=none] (22) at (-1, -0.5) {};
		\node [style=none] (23) at (1.25, -0.5) {$U_i$};
		\node [style=none] (25) at (-2.5, 1.75) {};
		\node [style=none] (26) at (-0.5, 0) {};
		\node [style=none] (28) at (-0.25, -1.25) {};
		\node [style=none] (29) at (-3, 2.25) {Backward staples};
		\node [style=none] (30) at (2.25, 1.25) {};
		\node [style=none] (31) at (3.5, 0.75) {};
		\node [style=none] (32) at (4, 2.25) {};
		\node [style=none] (33) at (4, 2.75) {Forward staples};
	\end{pgfonlayer}
	\begin{pgfonlayer}{edgelayer}
		\draw [style=new edge style 2] (1.center) to (2.center);
		\draw [style=new edge style 1] (0.center) to (7.center);
		\draw [style=new edge style 1] (6.center) to (5.center);
		\draw [style=new edge style 1] (2.center) to (3.center);
		\draw [style=new edge style 3] (5.center) to (4.center);
		\draw [style=new edge style 3] (3.center) to (4.center);
		\draw [style=new edge style 3] (7.center) to (6.center);
		\draw [style=new edge style 6] (0.center) to (19.center);
		\draw [style=new edge style 4] (19.center) to (20.center);
		\draw [style=new edge style 6] (20.center) to (21.center);
		\draw [style=new edge style 6] (2.center) to (17.center);
		\draw [style=new edge style 4] (21.center) to (18.center);
		\draw [style=new edge style 4] (17.center) to (18.center);
		\draw [style=new edge style 0] (0.center) to (1.center);
		\draw [style=new edge style 11] (11.center) to (12.center);
		\draw [style=new edge style 11] (9.center) to (8.center);
		\draw [style=new edge style 11] (9.center) to (10.center);
		\draw [style=new edge style 7] (8.center) to (0.center);
		\draw [style=new edge style 7] (10.center) to (11.center);
		\draw [style=new edge style 7] (12.center) to (2.center);
		\draw [style=new edge style 11] (14.center) to (22.center);
		\draw [style=new edge style 11] (14.center) to (13.center);
		\draw [style=new edge style 11] (16.center) to (15.center);
		\draw [style=new edge style 7] (22.center) to (16.center);
		\draw [style=new edge style 7] (13.center) to (0.center);
		\draw [style=new edge style 7] (15.center) to (2.center);
		\draw [style=new edge style 9] (25.center) to (26.center);
		\draw [style=new edge style 9] (25.center) to (28.center);
		\draw [style=new edge style 9] (32.center) to (31.center);
		\draw [style=new edge style 9] (32.center) to (30.center);
	\end{pgfonlayer}
\end{tikzpicture}}
		\caption{A link $U_i$ (in black) and four of its surrounding staples. Staples in blue are backward staples (i.e. of type $U_1^\dagger U_2^\dagger U_3$) and forward staples (i.e. of type $U_1 U_2^\dagger U_3^\dagger$) are depicted in red. Each staple (or each plaquette if we also count the link $U_i$) accounts for an independent factor of the factorized Metropolis filter (cf. section \ref{sec:fact}).}
		\label{fig:lattice_staple}
	\end{figure}
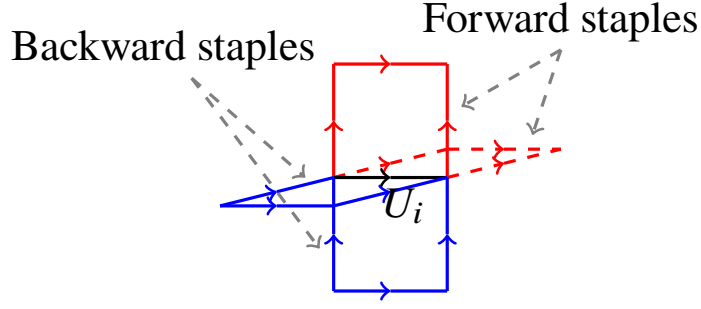

	Figure \ref{fig:lattice_staple} provides a geometric representation of the staples. Thus each link interacts only with other links in $2(d-1)$ surrounding plaquettes: $V^j_{x,\mu}$ is the staple matrix corresponding to the $j$-th plaquette attached to the link $U_{x,\mu}$. Such a factorization of the action into local interaction terms plays a central role in the construction of update algorithms. In particular, we can decompose the Boltzmann weight into contributions that correspond to individual plaquettes, a property that will be exploited in the formulation of the Event-Chain Monte Carlo dynamics described in next sections.
	
	\section{Event-Chain Monte Carlo}\label{sec:ecmc}
	ECMC is a continuous-time and rejection-free Markov Chain Monte Carlo scheme. Its implementation takes place in several steps. The appropriate formalism to define an Event-Chain algorithm is the Piecewise Deterministic Markov Process (PDMP) formalism \cite{monemvassitisPDMPCharacterisationEventchain2023} presented in section \ref{sec:pdmp}, but for the sake of pedagogy we present in this section the algorithm using an infinitesimal discrete Markov Chain approximation.
	
	\subsection{Factorization of the Metropolis filter}\label{sec:fact}
	First, the variation of action during the update of an element of the gauge configuration, in our case a link, needs to be separated in a sum of factors depending on the surrounding degrees of freedom of the system. This allows us to factorize the Metropolis filter in a product of independent factors.
	
	Let us denote from now on the links by a single index $i = 1,2,...,N_l$ with $N_l$ the number of links. In the case of Wilson action, eq. \eqref{eq:deltas} and \eqref{eq:staples} yields a decomposition
	\begin{equation}\label{eq:decomps}
		\Delta S(U_i\rightarrow U_i') = \sum_{j=1}^{2(d-1)} \Delta S_j(U_i\rightarrow U_i')
	\end{equation}
	with
	\begin{equation}
		\Delta S_j(U_i\rightarrow U_i') = -\frac{\beta}{N} \Re\Tr{(U_i'-U_i)V_i^j}.
	\end{equation}
	The regular Metropolis filter \cite{metropolisMonteCarloMethod1949} is defined by 
	\begin{equation}
		p(U_i\rightarrow U_i') = \min \left( 1, \ee^{-\Delta S(U_i\rightarrow U_i')} \right).
	\end{equation}
	Using the decomposition \eqref{eq:decomps} we write the factorized Metropolis filter \cite{krauthEventchainMonteCarlo2021} instead defined by
	\begin{equation}\label{eq:pfact}
		\pfact(U_i\rightarrow U_i') = \prod_{j=1}^{2(d-1)} \min\left(1, \ee^{-\Delta S_j(U_i\rightarrow U_i')}\right).
	\end{equation}
	Although the factorized Metropolis filter yields a smaller acceptance probability than the conventional Metropolis criterion, we can express the interaction of a link with its neighboring plaquettes as a sum of independent factors. This property is crucial for ECMC dynamics as it allows each plaquette contribution to be treated as an independent interaction term.
	
	\subsection{Updates and parametrization of SU(\texorpdfstring{$\bm{N}$}{N})}
	After factorizing the Metropolis filter, we need a way to explore SU($N$) during the link updates. We choose the XY-Embedding framework \cite{manaMultiGridMonteCarlo1997} as it can be easily generalized to any $N$. We will see in subsection \ref{sec:event} that it also makes the event generation part of the algorithm analytical and efficient.
	
	Following \cite{manaMultiGridMonteCarlo1997}, we update continuously the selected link variable through left multiplication parametrized as
	\begin{equation}
		U_i \rightarrow U_i(\theta) = R \ee^{\iu\epsilon \theta T} R^\dagger U_i ,\label{eq:update}
	\end{equation}
	where
	\begin{equation}
		\iu T = \mathrm{diag}(\iu,-\iu,0,\ldots,0)
	\end{equation}
	is a generator of $\mathfrak{su}(N)$ algebra, that are anti-hermitian traceless $N\times N$ matrices, $\theta$ is the evolution angle of the continuous update, $R$ a SU($N$) matrix chosen randomly according to the Haar measure and $\epsilon\in\{-1,1\}$ the direction of the update. Regular resampling of $R$ assures the ergodic exploration of SU($N$).
	
	The difference in action caused by an update of $\theta$ on the link $i$ reads
	\begin{align}
		\Delta S_{i, R, \epsilon}(\theta) = \sum_{j=1}^{2(d-1)} -\frac{\beta}{N}\Re \Tr{R\ee^{\iu \theta T}R^\dagger U_i V_i^j }.
	\end{align}
	In particular, this continuous updating process can be expressed as the limit of successive infinitesimal updates of angle $\dd\theta$, each accepted with the factorized Metropolis probability
	\begin{equation}
		\pfact(U_i \rightarrow U_i(\dd\theta)) = \prod_{j=1}^{2(d-1)} \min\left(1, \ee^{-\partial_\theta S^j_{i,R,\epsilon}(0) \dd\theta}\right) = \prod_{j=1}^{2(d-1)} \ee^{-\pll{\partial_\theta S^j_{i,R,\epsilon}(0)}\dd\theta},
	\end{equation}
	where each factor corresponds to the contribution of a single plaquette that includes the link $i$ and $\pll{x} = \max(0,x)$. The associated action increments at order $\dd\theta$ are
	\begin{equation}
		\partial_\theta S^j_{i,R,\epsilon}(0) = \epsilon\frac{\beta}{N}\Im\Tr{RT R^\dagger U_i V_i^{j} }.
	\end{equation}
	The Event-Chain is the infinitesimal limit of the following Markov Chain algorithm : we apply successive infinitesimal updates of angle $\dd\theta$ on the chosen link $i$ as long as all the factors of \eqref{eq:pfact} accept the update.
	The evolution continues until an \emph{lifting event} (or \emph{lift}) occurs. Lifting events arise from the rejection of the infinitesimal update by one of the factors of the acceptance probability.
	A lift proceeds as follows : when a staple $V^j_i$ rejects the update, we stop updating the link $i$ and start updating a link $k$ belonging to the staple $V^j_i$ instead. When $\dd\theta \rightarrow 0$, we can show \cite{michelGeneralizedEventchainMonte2014} that the probability of two factors or more rejecting an update goes to zero. Therefore the rejecting factor is uniquely determined.
	
	During a lift we can also change the values of $\epsilon$ and $R$, but in order to ensure convergence towards the right equilibrium distribution \eqref{eq:pi} we need to make sure that our algorithm respects global balance. In case of SU($N$) Yang-Mills lattice field theory regularized by Wilson action, we found two possible lift rules fulfilling global balance. We detail them in subsections \ref{sec:rev} and \ref{sec:irrev}. 
	
	We also need to define a total displacement angle $\theta_{\text{refresh}}$ upon which we break the chain and choose randomly with uniform probability a new link and a new $R$ matrix to ensure ergodicity of the algorithm.
	
	Sampling of gauge configuration happens after a fixed total displacement angle $\theta_{\text{sample}}>\theta_{\text{refresh}}$. Setting a fixed total displacement angle for sampling ensures that we do not sample at events, as this would introduce a bias easily seen in the values of observables.

	\subsection{Global balance and lifted state space}\label{sec:gb}
	
	The global balance equation is the main requirement to ensure that a Markov Chain algorithm converges towards the right probability distribution $\pi$ on a configuration space $\Omega$. In case of ECMC algorithms, we consider instead an extended version of the configuration space $\Omega$ called the \emph{lifted state space}. For SU($N$) lattice gauge theory on a 4 dimensional lattice of $N_l$ links, we consider the lift variables
	\begin{equation}
		(i, R, \epsilon) \in \mathcal{V} = \llbracket1,N_l\rrbracket \times \mathrm{SU}(N) \times \{-1,1\},
	\end{equation}
	which denote respectively the index of the link currently being updated, the rotation applied on the $\mathfrak{su}(N)$ generator used in the XY-embedding updates \eqref{eq:update} and the direction of the update.
	A configuration in the lifted state space is then characterized by the tuple
	\begin{equation}
		W = (U, i, R, \epsilon) \in \text{SU}(N)^{N_l}\times\mathcal{V},
	\end{equation}
	with $U \in \Omega \coloneq \text{SU}(N)^{N_l}$ the gauge configuration. Denoting $\mu$ the product of uniform probability distributions on the lift variables of space $\mathcal{V}$, the target probability distribution on the lifted state space is $\pi\otimes\mu$.
	
	Given a configuration space (in our case $\Omega \times \mathcal{V}$) and a transition operator $T$, the global balance equation states that for any lifted configuration, the sum of the probability flows to enter this state during the chain, so-called incoming probability flows, must equate the sum of the probabilities of departing from this state during the chain, so-called outgoing probability flows. Hence we write for any $W\in \Omega\times\mathcal{V}$
	
	\begin{equation}
		\sum_{W'\in \Omega\times\mathcal{V}}\pi\otimes\mu(W)T(W\rightarrow W') = \pi\otimes\mu(W) = \sum_{W'\in \Omega\times\mathcal{V}}\pi(W')T(W'\rightarrow W).\label{eq:gb1}
	\end{equation}
	
\begin{figure}[htbp]
		\centering
		\begin{subfigure}[b]{0.3\linewidth}
			\centering
			\resizebox{\linewidth}{!}{\begin{tikzpicture}
	\begin{pgfonlayer}{nodelayer}
		\node [style=new style 3] (0) at (0, 0) {};
		\node [style=new style 0] (1) at (-1.5, 1.5) {};
		\node [style=new style 0] (2) at (-2, 0) {};
		\node [style=new style 0] (3) at (-1.5, -1.5) {};
		\node [style=new style 0] (4) at (0, -2) {};
		\node [style=new style 0] (5) at (1.5, -1.5) {};
		\node [style=new style 0] (6) at (2, 0) {};
		\node [style=new style 0] (7) at (1.5, 1.5) {};
		\node [style=new style 0] (8) at (0, 2) {};
	\end{pgfonlayer}
	\begin{pgfonlayer}{edgelayer}
		\draw [style=new edge style 10, bend left=15] (0) to (2);
		\draw [style=new edge style 10, bend left=15] (0) to (4);
		\draw [style=new edge style 10, bend left=15] (0) to (6);
		\draw [style=new edge style 10, bend left=15] (0) to (7);
		\draw [style=new edge style 8, bend left=15] (7) to (0);
		\draw [style=new edge style 8, bend left=15] (6) to (0);
		\draw [style=new edge style 8, bend left=15] (4) to (0);
		\draw [style=new edge style 8, bend left=15] (2) to (0);
	\end{pgfonlayer}
\end{tikzpicture}}
			\caption{Detailed}
			\label{fig:gbdetailled}
		\end{subfigure}
		\hfill 
		\begin{subfigure}[b]{0.3\textwidth}
			\centering
			\resizebox{\linewidth}{!}{\begin{tikzpicture}
	\begin{pgfonlayer}{nodelayer}
		\node [style=new style 3] (0) at (0, 0) {};
		\node [style=new style 0] (1) at (-1.5, 1.5) {};
		\node [style=new style 0] (2) at (-2, 0) {};
		\node [style=new style 0] (3) at (-1.5, -1.5) {};
		\node [style=new style 0] (4) at (0, -2) {};
		\node [style=new style 0] (5) at (1.5, -1.5) {};
		\node [style=new style 0] (6) at (2, 0) {};
		\node [style=new style 0] (7) at (1.5, 1.5) {};
		\node [style=new style 0] (8) at (0, 2) {};
	\end{pgfonlayer}
	\begin{pgfonlayer}{edgelayer}
		\draw [style=new edge style 10, bend left=15] (0) to (2);
		\draw [style=new edge style 10, bend left=15] (0) to (4);
		\draw [style=new edge style 10, bend left=15] (0) to (6);
		\draw [style=new edge style 10] (0) to (7);
		\draw [style=new edge style 8, bend left=15] (6) to (0);
		\draw [style=new edge style 8, bend left=15] (4) to (0);
		\draw [style=new edge style 8, bend left=15] (2) to (0);
		\draw [style=new edge style 8] (5) to (0);
		\draw [style=new edge style 10] (0) to (1);
	\end{pgfonlayer}
\end{tikzpicture}}
			\caption{Partially irreversible}
			\label{fig:gbrev}
		\end{subfigure}
		\hfill
		\begin{subfigure}[b]{0.3\textwidth}
			\centering
			\resizebox{\linewidth}{!}{\begin{tikzpicture}
	\begin{pgfonlayer}{nodelayer}
		\node [style=new style 3] (0) at (0, 0) {};
		\node [style=new style 0] (1) at (-1.5, 1.5) {};
		\node [style=new style 0] (2) at (-2, 0) {};
		\node [style=new style 0] (3) at (-1.5, -1.5) {};
		\node [style=new style 0] (4) at (0, -2) {};
		\node [style=new style 0] (5) at (1.5, -1.5) {};
		\node [style=new style 0] (6) at (2, 0) {};
		\node [style=new style 0] (7) at (1.5, 1.5) {};
		\node [style=new style 0] (8) at (0, 2) {};
	\end{pgfonlayer}
	\begin{pgfonlayer}{edgelayer}
		\draw [style=new edge style 8] (3) to (0);
		\draw [style=new edge style 8] (5) to (0);
		\draw [style=new edge style 10] (0) to (2);
		\draw [style=new edge style 10] (0) to (4);
		\draw [style=new edge style 10] (0) to (6);
		\draw [style=new edge style 10] (0) to (7);
	\end{pgfonlayer}
\end{tikzpicture}}
			\caption{Maximally irreversible}
			\label{fig:gbirrev}
		\end{subfigure}
		\caption{Illustration of detailed balance (\ref{fig:gbdetailled}), partially irreversible global balance (\ref{fig:gbrev}) and maximally irreversible global balance (\ref{fig:gbirrev}). We depict the incoming (in blue, dashed) and outgoing (in green) probability flows into a configuration of the considered state space $\Omega$.}
		\label{fig:gb}
	\end{figure}
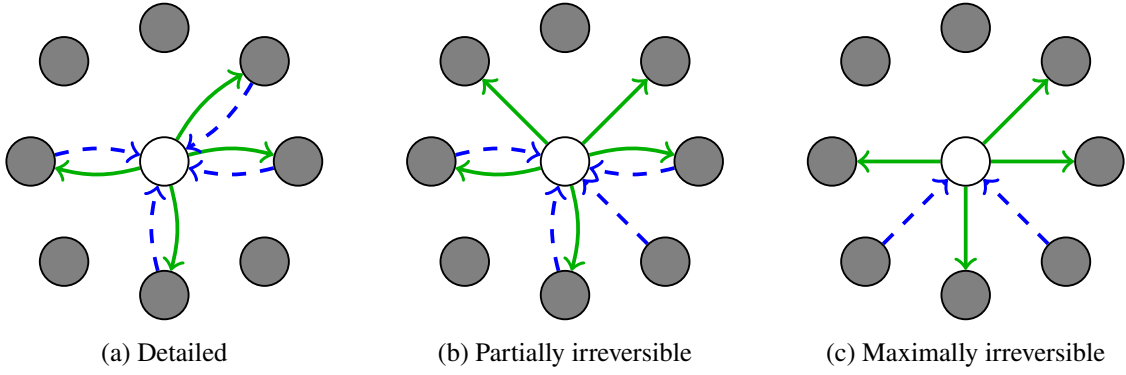

	Usual MCMC algorithms enforce a stronger version called \emph{detailed} balance on $\Omega$ where the equality \eqref{eq:gb1} is term-wise. The purpose of ECMC is to fulfill global balance without the detailed version in the lifted configuration space. It ensures a ballistic exploration of the configuration space instead of a diffusive one.
	Figure \ref{fig:gb} depicts the different ways algorithms fulfill \eqref{eq:gb1}. Algorithms such as Metropolis, Heatbath and HMC verify the detailed balance, while both algorithms presented in section \ref{sec:rev} and \ref{sec:irrev} preserve a maximally irreversible global balance in the lifted state space.
	We show convergence of our ECMC algorithms using the rigorous PDMP formalism in section \ref{sec:invariance}.

	\subsection{Event-driven approach}\label{sec:event}
	The Event-Chain algorithm generates a continuous-time Markov process in the lifted configuration space but that can be exactly simulated by an event-driven approach \cite{bernardEventchainMonteCarlo2009, michelGeneralizedEventchainMonte2014, petersEventChainMonteCarlo2026} yielding the discrete sequences of events. Event angles $\theta_{\mathrm{event}}^j$ are sampled for each farctor $j$ and represent a realization of the Event-Chain where the $j$-th factor accepted the update of the link up to $\theta_{\text{event}}^j$ and then rejected the update. Factorization of the Metropolis filter allows us to treat each factor independently. We can then update in a single step the link with an angle $\theta_{\text{event}} = \min_{j = 1,...,2(d-1)}\theta_{\text{event}}^j$, i.e. up to the point where one of the factors rejects the update. This update of the chosen link $i$ between two lifts, in a purely deterministic way and in a single step, does achieve a substantial speed-up with respect to usual MCMC algorithms with infinitesimal updates such as Metropolis algorithm. This allows for a ballistic exploration of the configuration space, by opposition to the diffusive exploration of MCMC algorithms designed with local updates.
	
	We can sample event angles for each member $j$ of the factorized Metropolis filter by solving for $\theta^j_{\text{event}}$ with $u$ drawn uniformly in $(0,1]$ \cite{bernardEventchainMonteCarlo2009, michelGeneralizedEventchainMonte2014, michelEventchainMonteCarlo2015, petersEventChainMonteCarlo2026}
	\begin{equation}
		-\ln(u) = \int_0^{\theta^j_{\text{event}}} \pll{\partial_\theta S_{i,R,\epsilon}^j(\theta)} \dd \theta. \label{eq:event}
	\end{equation}
	
	The equation \eqref{eq:event} can be retrieved in the infinitesimal approximation of the Event-chain as detailed in \cite{michelEventchainMonteCarlo2015}.
	Using the XY-Embedding, updating the links with \eqref{eq:update} we obtain an embedded Wilson action (we write only the $\theta$-dependent part)
	\begin{align}
		S_{i,R,\epsilon}^j (\theta) = -\frac{\beta}{N} (A^j_i \cos(\theta) + B^j_i \epsilon \sin(\theta)) = -\frac{\beta}{N} C^j_i \cos(\theta - \theta^j), \label{eq:embd_action}
	\end{align}
	with
	\begin{align}
		A^j_i &= \Re[(R^\dagger U_i V^j_i R)_{11}+(R^\dagger U_i V^j_i R)_{22}],\\
		B^j_i &= \Im[(R^\dagger U_i V^j_i R)_{11}-(R^\dagger U_i V^j_i R)_{22}],\\
		C^j_i &= \sqrt{(A^j_i)^2 + (B^j_i)^2},\\
		\theta^j_i &= \arctan(\epsilon B^j_i/A^j_i).
	\end{align}
	The analytical resolution of \eqref{eq:event} made possible with the form of \eqref{eq:embd_action} ensures an efficient sampling of event angles and a continuous exploration of SU($N$) for each link visited by the chain.

	\subsection{Reflective ECMC lift rule}\label{sec:rev}
	
	After updating the link up to the angle $\theta_{\text{event}}$, we then proceed to \emph{lift} to another state of the lifted state space by changing the lift variables $(i,\epsilon)\rightarrow (k,\hat\epsilon)$ with $k$ a link of the plaquette $P^j$ responsible for $\theta_{\text{event}}$. We lift towards a link $k$ of this plaquette with probability
	\begin{align}\label{eq:probrev}
		\lambda_{(i,\epsilon)\rightarrow (k,\hat\epsilon)} = \frac{|\partial_\theta S^j_{k,R}(0)|}{\sum_{l \in P^j} |\partial_\theta S_{l,R}^j(0)|}\delta(\hat\epsilon + \text{sign}(\partial_\theta S^j_{k,R}(0))),
	\end{align}
	using the notation $\partial_\theta S^j_{k,R} = \frac{\beta}{N}\Im\Tr{T R^\dagger U_i V_i^{j} R}\dd\theta$. This lift rule is a variant of \cite{harlandEventchainMonteCarlo2017} where we enable lifts to the same link $i$ with opposite update direction $\epsilon$. This addition is mandatory for the conservation of global balance (cf. section \ref{sec:invariance}) as the system does not exhibit the same symmetries as \cite{harlandEventchainMonteCarlo2017}. Indeed, the non-Abelian nature of SU($N$) yields
	\begin{equation}
		\sum_{i\in P^j} \partial_\theta S^j_i(0) \neq 0.
	\end{equation}
	This lifting rule can therefore induce backtracking. However, this lift rule is general and not tied to the Wilson action.
	
	Then we repeat the same steps with the new link $k$ selected and $\hat\epsilon = -\text{sign}(\partial_\theta S_{k,R}^j(0))$.
	
	\subsection{Forward ECMC lift rule} \label{sec:irrev}
	
	The reflective lift rule previously discussed generates backtracking due to the non-Abelian nature of SU($N$). It is however possible to exploit an intrinsic symmetry of the considered system, namely the invariance of the Wilson action under charge conjugation and the geometry of the plaquette, i.e. a closed path. This enables the definition of an alternative lifting rule which completely suppresses backtracking.  When an event arises, we choose one of the links $k$ of the plaquette $P^j$ involved in the event, $k \neq i$, with uniform probability $1/3$ and perform the lift
	\begin{align}
		(U,i,R,\epsilon) \rightarrow (U,k,R_k,\epsilon_k).
	\end{align}
	$(R_k,\epsilon_k)$ depends wether  $P^j$ is a forward or a backward plaquette (cf. Fig. \ref{fig:plaquette}) and on the position of link $k$ within this plaquette.
	
	Let's index the current link of the chain with $i=1$ and the other links of the rejecting plaquette with subsequent integers such that $P^j = U_1U_2U_3^\dagger U_4^\dagger$ if the plaquette is forward for $U_1$ and $P^j = U_1 U_2^\dagger U_3^\dagger U_4$ if the plaquette is backward. The corresponding lifting rules are detailed in table \ref{tab:lift}.
	
	\begin{table}[t]
		\centering
		\renewcommand{\arraystretch}{1.2} 
		\begin{tabular}{ccccc}
			& \multicolumn{2}{c}{Forward} & \multicolumn{2}{c}{Backward} \\
			\cmidrule(lr){2-3} \cmidrule(lr){4-5}
			$k$  & $R_k$ & $\epsilon_k$ & $R_k$ & $\epsilon_k$  \\
			\midrule
			$2$  & $U_1^\dagger R$ & $-\epsilon$  & $U_3^\dagger U_4 R$ & $\epsilon$\\
			$3$  &  $U_4^\dagger R$&   $\epsilon$ & $U_4 R$ & $\epsilon$ \\
			$4$   &  $R$           &   $\epsilon$ &  $U_4 R$ & $-\epsilon$ \\
			\bottomrule
		\end{tabular}
		\caption{Lifting rules for $R_k$ and $\epsilon_k$ depending on the plaquette type and the link $k$ position.}
		\label{tab:lift}
	\end{table}
	Thanks to the suppression of backtracking, this irreversible dynamic induces an extra speed-up with respect to the reflective dynamics shown in Section \ref{sec:results}. The frequent updates of $R$ also enables a more efficient exploration of SU($N$).  The proof of the conservation of global balance in this framework is detailed in section \ref{sec:invariance} and relies on the fact that for each $(i,k)$ in plaquette $P^j$, a lift from $(i,R,\epsilon)$ to $(k,R_k, \epsilon_k)$ verifies $\partial_\theta S^j_{k, R_k, \epsilon_k}(0) = -\partial_\theta S^j_{i, R, \epsilon}(0)$.
	
	This lifting rule is action specific, but an Event-Chain algorithm with similar irreversible lift rules can be implemented with improved action containing traces of closed loops such as the L\"uscher-Weisz \cite{luscherOnshellImprovedLattice1985} or the Iwasaki \cite{iwasakiRenormalizationGroupAnalysis2011} actions in case of quenched gauge lattice QCD. Those actions are also factorizable and for a lift from link $i$ towards a link $k$ of a factor $j$, namely a trace of a closed loop containing $i$ and $k$, we can always define $(R_k, \epsilon_k)$ such that $\partial_\theta S^j_{k, R_k, \epsilon_k}(0) = -\partial_\theta S^j_{i, R, \epsilon}(0)$, using hermitian conjugation and cyclic permutations inside the trace.
	
	\subsection{Summary}
	To summarize the Event-Chain algorithm with XY-embedding:
	
	\begin{enumerate}
		\item Define a total displacement length $\Thetas$ upon which the sample is saved, and a total displacement length $\Thetar \ll \Thetas$ upon which we refresh the matrix $R$, the direction $\epsilon$ and the link to update. 
		\item Initialize total displacement counters $\thetar$ and $\thetas$ to zero.
		\item Choose a link $i$, a matrix $R\in$ SU($N$) and $\epsilon\in \{-1,1\}$ randomly with uniform probability.
		\item Generate a reject angle $\theta^j_{\text{event}}$ for each of the plaquettes $P^j$, $j=1,2,...,2(d-1)$ attached to the chosen link. The smallest angle $\theta_{\text{event}}^{j} = \theta_{\text{event}}$ will be the angle chosen for the update. (it means that the corresponding plaquette $j$ was the first one to reject a move) 
		\item Update the link $i$ according to \eqref{eq:update} with $\theta = \min(\theta_{\text{event}}, \Thetas - \thetas, \Thetar-\thetar)$.
		\begin{enumerate}
			\item if $\theta = \thetae$, update the total displacement counters $\thetar \to \thetar + \theta$ and $\thetas \to \thetas+ \theta$ and go to step 6.
			\item if $\theta = \Thetar-\thetar$, update $\thetar \to 0,$ $\thetas \to \thetas + \theta$ and go to step 3.
			\item if $\theta = \Thetas - \thetas$, save the configuration as a sample, set $\thetas \to 0$, $\thetar \to \thetar + \theta$. Then redo step 5 with $\thetae \to \thetae-\theta$.
		\end{enumerate}
		
		\item Lift by choosing one of the links $k$ of $P^j$ :
		\begin{itemize}
			\item with uniform probability if we use the forward lift rule. Also change $(R,\epsilon)$ according to Table \ref{tab:lift},
			\item with probability \eqref{eq:probrev} if we use the reflective lift rule. Also change $\epsilon \rightarrow -\text{sign}(\partial_\theta S_{k,R}^j(0))$.
		\end{itemize}
		and repeat from step 4.
	\end{enumerate}
	
\begin{figure}[htbp]
		\centering
		\scalebox{1.2}{\begin{tikzpicture}
	\begin{pgfonlayer}{nodelayer}
		\node [style=new style 3] (0) at (-4, 0) {1};
		\node [style=new style 3] (1) at (-1.5, 0) {2};
		\node [style=new style 3] (2) at (1, 0) {3};
		\node [style=new style 3] (3) at (3.5, 0) {4};
		\node [style=new style 3] (4) at (6, 0) {5};
		\node [style=new style 3] (5) at (8.5, 0) {5b};
		\node [style=new style 3] (6) at (7, 2.5) {5a};
		\node [style=new style 3] (7) at (7, -2.5) {5c};
		\node [style=new style 3] (8) at (10.5, 0) {6};
	\end{pgfonlayer}
	\begin{pgfonlayer}{edgelayer}
		\draw [style=new edge style 0] (0) to (1);
		\draw [style=new edge style 0] (1) to (2);
		\draw [style=new edge style 0] (2) to (3);
		\draw [style=new edge style 0] (3) to (4);
		\draw [style=new edge style 0] (4) to (6);
		\draw [style=new edge style 0] (4) to (5);
		\draw [style=new edge style 0] (4) to (7);
		\draw [style=new edge style 0] (6) to (8);
		\draw [style=new edge style 0, bend left=270, looseness=1.50] (8) to (3);
		\draw [style=new edge style 0, bend left=105, looseness=1.50] (5) to (2);
		\draw [style=new edge style 0, bend right=300, looseness=1.25] (7) to (4);
	\end{pgfonlayer}
\end{tikzpicture}}
		\caption{Execution flow of the ECMC algorithm. The numbered nodes correspond to the steps outlined in the summary of the algorithm.}
		\label{fig:algo}
	\end{figure}
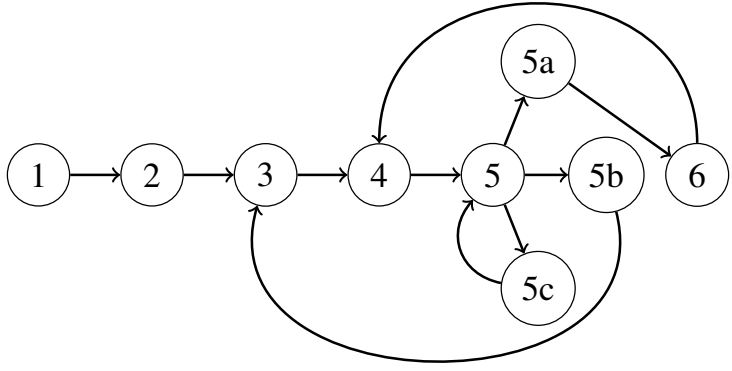

	Figure \ref{fig:algo} depicts the flow of the algorithm. The refreshment steps straightforwardly ensures the ergodicity of such ECMC algorithm. It is also possible to split the refresh of each of the lift variables $(i, R,\epsilon)$ by defining a counter $\theta$ and a total displacement angle $\Theta$ for each, but the gain in autocorrelation time is minor with respect to the large increase of the cost of fine tuning those parameters.
	
	\section{PDMP formalism}\label{sec:pdmp}
	
	The efficient formalism to define an ECMC algorithm is the Piecewise Deterministic Markov Process (PDMP) framework \cite{monemvassitisPDMPCharacterisationEventchain2023, guyonNecessarySufficientSymmetries2025}. Starting from any lifted state $(U,v) \in \Omega \times \mathcal{V}$, an Event-Chain can be seen as a process $\{U(\theta), v(\theta)\}_{\theta \geq 0}$ on the lifted configuration space $\Omega\times \mathcal{V}$. $U(\theta)=(U_1(\theta), U_2(\theta), ..., U_{N_l}(\theta)) \in \Omega$ denotes the evolution of a gauge configuration with $N_l$ links, starting from $U(0)\equiv U$, during the ECMC process and $v(\theta)=(i(\theta),R(\theta),\epsilon(\theta))\in \mathcal{V}$ the evolution of the lift variables, starting from $V(0) \equiv V$, during the ECMC process. We can then characterize an Event-Chain algorithm with an infinitesimal generator $\mathcal{A}$. This framework allows a definition of the ECMC process directly as a continuous-time process, without resorting to the infinitesimal Markov Chain limit, and yields a simple condition that the stationary distribution must satisfy.
	
	\subsection{Definitions}
	
	Following \cite{monemvassitisPDMPCharacterisationEventchain2023}, we define the infinitesimal generator $\mathcal{A}$ by
	\begin{equation}
		\mathcal{A}f = \lim_{\theta\rightarrow 0} \frac{\mathbb{E}_{U,v}[f(U(\theta),v(\theta)) - f(U,v)]}{\theta}.
	\end{equation} 
	
	The ECMC process can be decomposed in two parts : a deterministic flow $\phi$ corresponding to the macroscopic updates between two lifts and stochastic events leading to lift and refreshment events affecting the lift variables.
	
	In our case, the deterministic flow corresponds to the updates \eqref{eq:update} of the chosen link. Thus we define for $\theta\geq0$
	\begin{equation}
		(\dot U, \dot v) = \phi(U,v) = (\phi_1(U,v), \phi_2(U,v)),
	\end{equation}
	with the dot denoting derivation with respect to $\theta$ and
	\begin{equation}\label{eq:flot}
		\begin{aligned}
			\phi_1 (U,v) &= (0, 0, ... , \underbrace{\epsilon R \iu T R^\dagger U_i}_{\text{coordinate }i}, 0,..., 0) \quad \text{ if } v = (i,R,\epsilon),\\
			\phi_2(U,v) &= 0.
		\end{aligned}
	\end{equation}

	The events are characterized by the event rate $\lambda : \Omega\times \mathcal{V} \rightarrow \mathbb{R}_+$ and a Markov kernel $Q$ defined on $\Omega\times \mathcal{V} \times \mathcal{B}(\Omega\times\mathcal{V})$ which sums up the different lifting rules. We voluntarily omit the boundary kernel $Q^b$ describing the refresh process for the sake of clarity as it does not intervene in the invariance computation \cite{monemvassitisPDMPCharacterisationEventchain2023}.
	
	The general form of the generator $\mathcal{A}$ for an Event-Chain is \cite{guyonNecessarySufficientSymmetries2025}
	\begin{equation}\label{eq:gen}
		\mathcal{A}f(U,v) = \langle \phi, \nabla_{U,v} f(U,v) \rangle + \lambda(U,v) \int_{\Omega\times\mathcal{V}} (f(U',v')- f(U,v))Q[(U,v), (\dd U', \dd v')] ,
	\end{equation}
	with the first term coding for the deterministic drift and the second term for changes due to the stochastic events determined by the Markov kernel $Q$. 
	
	In our case, for a lifted state $(U,v)$, the event rate is the sum of event rates $\lambda^j$ due to all the plaquettes $P^j$ attached to link $i$. We can write, with $v=(i,R,\epsilon)$,
	
	\begin{equation}\label{eq:facteventrate}
		\lambda(U,v) = \sum_{j=1}^{2(d-1)} \lambda^j(U,v) = \sum_{j=1}^{2(d-1)} \pll{\partial_\theta S^j_{v}(0)} = \sum_{j=1}^{2(d-1)} \pll{\langle \phi, \nabla_{U_i} S^j_{v}\rangle}.
	\end{equation}
	with $\langle A,B \rangle = \Re\Tr{A^\dagger B}$ the usual scalar product on SU($N$) and $\nabla_{U_i} S^j_v$ denotes the gradient of the action with respect to the matrix elements of $U_i$.  Formally, it corresponds to the Lie derivative of the action along the algorithmic time $\theta$. Note that the other links of configuration $U$ have no impact in the scalar product because of \eqref{eq:flot}.
	
	The Markov kernel $Q$ can also be decomposed in a weighted sum of Markov $Q^j$ coding for the lifts in plaquette $P^j$. We have
	\begin{equation}
		\begin{aligned}
			Q((U,v), (\dd U', \dd v')) &= \sum_{j=0}^{2(d-1)} \underbrace{\frac{\lambda^j(U,v)}{\lambda(U,v)}}_{\substack{\text{Probability for } P^j \\ \text{ to trigger an event}}} Q^j[(U,v), (\dd U', \dd v')]
		\end{aligned}
	\end{equation}
	We can then write \eqref{eq:gen} as
	\begin{equation}
		\mathcal{A}f(U,v) = \langle \phi, \nabla_{U,v} f(U,v) \rangle +  \int_{\Omega\times\mathcal{V}} (f(U',v')- f(U,v))\sum_{j=1}^{2(d-1)}\lambda^j(U,v) Q^j[(U,v), (\dd U', \dd v')],
	\end{equation}
	
	The definition of $Q^j$ depends on the lifting scheme used. In case of reflective ECMC, we have
	\begin{equation}\label{eq:kernelrev}
		Q^j_{\text{reflective}}[(U, v), (\dd U', \dd v')] = \sum_{k\in P^j} \frac{|\partial_\theta S^j_{k,R}(0)|}{\sum_{l \in P^j} |\partial_\theta S_{l,R}^j(0)|} \delta_U(\dd U') \delta_{(k, R, -\text{sign}(\partial_\theta S^j_{k,R}(0)))}(\dd v'),
	\end{equation}
	and in the forward ECMC case we have
	
	\begin{equation}\label{eq:kernelirrev}
		Q^j_{\text{forward}}[(U,v), (\dd U', \dd v')] = \sum_{k\in P^j\setminus \{i\}} \frac{1}{3} \delta_U(\dd U') \delta_{(k,R_k,\epsilon_k)}(\dd v'),
	\end{equation}
	with $R_k, \epsilon_k$ defined in table \ref{tab:lift}.
	
	\subsection{Invariance of the target probability distribution}\label{sec:invariance}
	
	The invariance of the target probability distribution is expressed as continuous-time variant of the global balance equation \eqref{eq:gb1} written as
	\begin{equation}
		\int_{\Omega\times \mathcal{V}} \mathcal{A}f(X) \pi(X) \dd X = 0.
	\end{equation}
	
	Following the derivations of \cite{guyonNecessarySufficientSymmetries2025}, a sufficient condition for algorithms defined with factorized event rates (cf. \eqref{eq:facteventrate}) is written for each factor $j$ and any lifted state $(U,v)$
	\begin{equation}\label{eq:invariancesufficient}
		\pmm{\partial_\theta S^j_v(0)} \mu(v) = \int_{\mathcal{V}} \dd v' \mu(v') \pll{\partial_\theta S^j_{v'}(0)} Q^j[(U,v'), (U,v)],
	\end{equation}
	with $\pmm{x} = \max(0,-x)$ the negative part of $x$, such that $x=\pll{x} - \pmm{x}$.
	
	Condition \eqref{eq:invariancesufficient} dictates that any local dissipation caused by the deterministic drift must be perfectly counterbalanced by redistribution performed by the Markov kernels during lift events. Heuristically, the stochastic events act as probability regulators that perpetually replenish the target measure wherever the continuous deterministic flow tends to drain it. 
	
	Given the expressions \eqref{eq:kernelrev} and \eqref{eq:kernelirrev} of the Markov kernel used in both implementations, showing that this condition holds is straightforward. For the reflective ECMC kernel, by definition, $Q^j_{\text{reflective}}[(U,v'), (U,v)] \neq 0$ only if $v$ is such that updating from the lifted state $(U,v)$ decreases the action. Thus we have for the l.h.s. of \eqref{eq:invariancesufficient} $\pmm{\partial_\theta S^j_v(0)} = |\partial_\theta S^j_v(0)|$.
	
	Furthermore, denoting  $v = (i,R,\epsilon)$, we have
	\begin{equation}
		Q^j_{\text{reflective}}[(U,v'), (U,v)] \neq 0 \Leftrightarrow v'\in P^j \times \{ R \} \times \{-\text{sign}(\partial_\theta S^j_{i,R}(0))\} \coloneqq \mathcal{P}_v.
	\end{equation}
	
	This yields 
	
	\begin{equation}
		\begin{aligned}
			&\int_{\mathcal{V}} \dd v' \mu(v') \pll{\partial_\theta S^j_{v'}(0)} Q^j_{\text{reflective}}[(U,v'), (U,v)]\\
			&\quad=\sum_{v'\in \mathcal{P}_v} \mu(v') \pll{\partial_\theta S^j_{v'}(0)} \frac{|\partial_\theta S^j_{v}(0)|}{\sum_{l \in P^j} |\partial_\theta S_{l,R}^j(0)|}
		\end{aligned}
	\end{equation}
	
	We then have $\mu(v') = \mu(v)$, as the probability distribution is uniform on $\mathcal{V}$, and 
	$\sum_{l \in P^j} |\partial_\theta S_{l,R}^j(0)| = \sum_{v'\in \mathcal{P}_v} \pll{\partial_\theta S^j_{v'}(0)}$, ensuring the realization of condition \eqref{eq:invariancesufficient}.
	
	For the forward lift rule, the kernel is non zero if $v' \in \{(k, R_k, \epsilon_k), k \in P^j\setminus\{i\}\}\coloneqq \mathcal{P}_v$, which contains 3 elements (the 3 links of the corresponding staple). The selection rules of table \ref{tab:lift} ensure $\pll{\partial_\theta S^j_{v'}(0)} = \pmm{\partial_\theta S^j_v(0)}$. The r.h.s. of condition \eqref{eq:invariancesufficient} becomes
	\begin{equation}
		\begin{aligned}
			&\int_{\mathcal{V}} \dd v' \mu(v') \pll{\partial_\theta S^j_{v'}(0)} Q^j_{\text{forward}}[(U,v'), (U,v)]\\
			&\quad=\sum_{v'\in \mathcal{P}_v} \mu(v') \pll{\partial_\theta S^j_{v'}(0)} \left( \sum_{k\in P^j\setminus \{i\}} \frac{1}{3} \delta_{(k,R_k,\epsilon_k)}(v) \right)\\
			&\quad= \mu(v) \sum_{v'\in \mathcal{P}_v}  \frac{1}{3} \pmm{\partial_\theta S^j_v(0)}\\
			&\quad= \mu(v)\pmm{\partial_\theta S^j_v(0)}.
		\end{aligned}
	\end{equation}
	
	We have thus proven that the generators of reflective and forward ECMC algorithms leave the target probability distribution $\pi \otimes \mu$ invariant. Thanks to refreshment events, ergodicity is ensured \cite{monemvassitisPDMPCharacterisationEventchain2023} and therefore this shows the correct convergence of the algorithms in the lifted state space. 
	
	\section{Implementation for SU(3)}\label{sec:results}
	
	We can now generate configurations in pure gauge lattice QCD with Wilson action using both ECMC implementations and compare the measured value of observables to other sampling algorithms. In this proof-of-concept work, we use small lattice volumes of points to show correct thermalization and convergence towards the equilibrium distribution without the need for parallelization. Parallelization of ECMC for lattice QCD is not straightforward because of the stochastic nature of the lift process and it will be addressed in a forthcoming publication.
	
	The plaquette measurements have been obtained using both implementations of ECMC and quasi-Heatbath on $8^4$ lattices with periodic boundary conditions and different values of $\beta$. We define the mean plaquette of a gauge configuration of volume $V$ as
	\begin{equation}
		P = \frac{1}{18V} \sum_{x \in \Lambda} \Re\Tr{U_{x,\mu\nu}}
	\end{equation}
	with $\Lambda$ the set of lattice sites. Measurements of the mean plaquette were performed every sweep. In order to compare ECMC to sequential algorithms such as Heatbath or Metropolis, we need to choose $\thetas$ such that the number of events between two samples is equal to the number of links. This value depends on the size of the lattice and on the inverse bare coupling $\beta$. The values chosen are detailed in Table \ref{tab:thetas}.
	\begin{table}[t]
		\centering
		\begin{tabular}{cccccccccc}
			\toprule
			$\beta$ & 1 & 2 & 3 & 4 & 5 & 6 & 7 & 8 & 9 \\
			\midrule
			$\thetas$ & \num{32768}  & \num{16384} & \num{10649}  & \num{7700} & \num{6062} & \num{5242} & \num{4751} & \num{4423} & \num{4096} \\
			\bottomrule
		\end{tabular}
		\caption{Values of $\thetas$ required to define a sweep depending on $\beta$ for a $8^4$ lattice.}\label{tab:thetas}
	\end{table}
	High values of $\thetas$ for low $\beta$ are explained by the form of the factorized Metropolis filter \eqref{eq:pfact} and the computation of $\thetae$ \eqref{eq:event} which allows for larger update angles when the bare coupling is low.
	
	The analysis of the measures is performed with the $\Gamma$-method \cite{wolffMonteCarloErrors2004}. For a thermalized run yielding a finite number of values $(P_1, P_2,...,P_{N_s})$ of the mean plaquette, we define $C(\tau)$ the autocorrelation at lag $\tau$ as
	\begin{equation}
		C(\tau) = \frac{1}{N-\tau} \sum_{i=1}^{N-\tau} (P_i - \bar{P})(P_{i+\tau} - \bar{P})
	\end{equation}
	where $\bar{P} = \frac{1}{N} \sum_{i=1}^N P_i$ is the sample mean. The normalized autocorrelation function is then given by $\rho(\tau) = C(\tau)/C(0)$. The crucial quantity for the error estimation is the integrated autocorrelation time, defined as
	\begin{equation}
		\tau_{\text{int}} = \frac{1}{2} + \sum_{\tau=1}^{\infty} \rho(\tau)
	\end{equation}
	In practice, the sum is truncated at a self-consistent window $W$ to avoid the accumulation of noise from the tail of $\rho(\tau)$ where the signal-to-noise ratio is low. The variance of the sample mean is then correctly estimated by
	\begin{equation}
		\sigma_{\bar{P}}^2 = \frac{C(0)}{N} \times 2\tau_{\text{int}}
	\end{equation}
	This method accounts for the correlations between successive configurations in the Markov Chain, which otherwise would lead to a significant underestimation of the statistical uncertainties. 
	
	In Section \ref{sec:results_mean}, we report the average plaquette values  obtained across a range of $\beta$ for both hot and cold starts of the ECMC implementations. These results are compared against Heatbath benchmarks to show algorithmic convergence. Section \ref{sec:tau_int} presents the comparison of autocorrelation times both in sweeps and CPU time between both implementations of ECMC and Heatbath.
	
	\subsection{Convergence}\label{sec:results_mean}
	
\begin{figure}[t]
		\centering
		\includegraphics[width=0.8\linewidth]{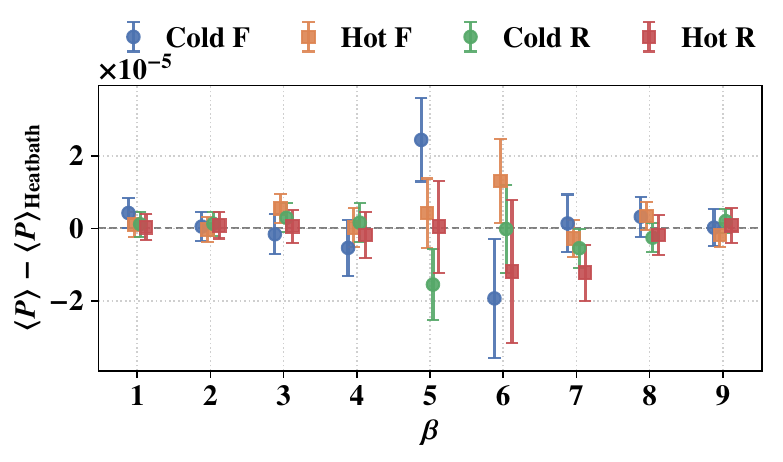}
		\caption{Comparison of residuals of the average plaquette with respect to Heatbath for reflective (R) and forward (F) ECMC implementations, with hot and cold starts.}\label{fig:plaquette}
	\end{figure}

	To assess convergence towards the target distribution $\pi$ for both ECMC implementations, we generate chains starting with a cold (every link to identity) and a hot (all links chosen at random in SU(3)) configuration. We also generate measurements using Heatbath. Comparison of the values of the average plaquette in all cases with Heatbath are shown in Figure \ref{fig:plaquette}. 
	
	We observe a very good agreement up to $\num{1e-5}$ with no visible bias, confirming a correct convergence. Tables that collect the values obtained and statistical consistency with respect to Heatbath are available in Appendix \ref{sec:details}.
	
	\subsection{Comparison of autocorrelation times}\label{sec:tau_int}
	
	Although the mean plaquette is a short-distance observable which has been shown to decouple from the slow modes of the theory \cite{schaeferCriticalSlowingError2011} and not subject to critical slowing down, the study of its autocorrelation can give us a first idea of the behavior of ECMC around the deconfinement phase transition with respect to other algorithms.
	\begin{figure}[t]
		\centering
		\includegraphics[width=\linewidth]{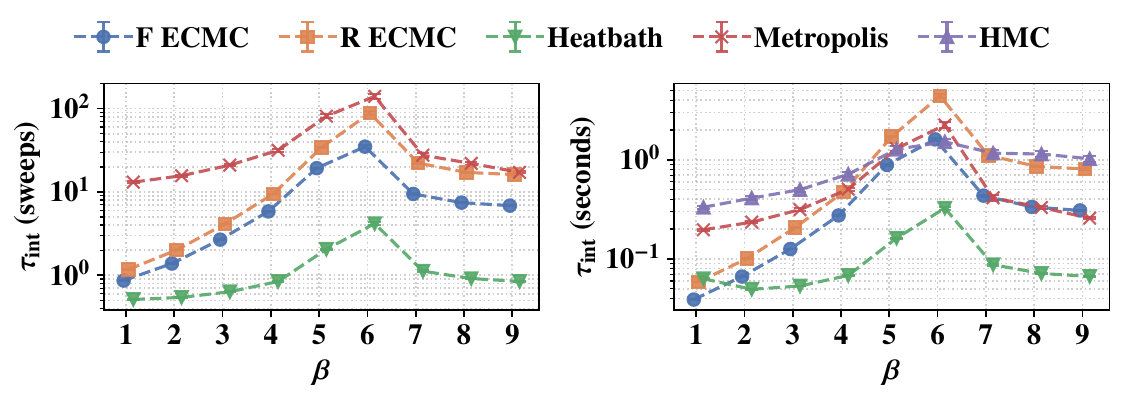}
		\caption{Comparison of autocorrelation times in sweeps (left) and in CPU time (right) as a function of $\beta$ between reflective (R) and forward (F) ECMC implementations, Heatbath, Metropolis and HMC.}\label{fig:tau_int}
	\end{figure}
	
	Figure \ref{fig:tau_int} illustrates the evolution of the integrated autocorrelation times as functions of $\beta$. It reveals a consistent behavior across all algorithms, with a pronounced peak near $\beta = 6$,  that is close to the deconfinement phase transition point $\beta_c(N_t=8) \sim 6.06$ \cite{boydThermodynamicsSU3Lattice1996a}. Both ECMC variants achieve superior sampling efficiency compared to the Metropolis algorithm in number of sweeps, though they do not reach the performance level of the Heatbath algorithm. When compared to Metropolis and HMC\footnote{We have set a trajectory lengh of molecular dynamics $\tau=3$ in our numerical test.} algorithms, forward ECMC achieves significantly lower CPU autocorrelation times away from the phase transition, while all three algorithms reach comparable efficiency in its vicinity. A direct comparison between the two ECMC implementations further reveals that the forward lifting rule yields more efficient decorrelation than its reflective counterpart, with a twofold speedup in the vicinity of the deconfinement transition. This improvement is consistent with the suppression of backtracking inherent to the forward dynamics, which enables a more thorough exploration of the configuration space.
	
	The performance gap with respect to Heatbath is not surprising, as the latter constitutes a highly optimized baseline: by sampling new link variables directly from the local conditional distribution, it achieves near-complete renewal of the local degrees of freedom at each update step. As such, Heatbath provides a natural lower bound for the autocorrelation time of local observables, against which the rejection-free yet potentially more ballistic dynamics of ECMC must be assessed.
	
	It should nonetheless be emphasized that the primary advantage of ECMC is expected to manifest in the sampling of long-range observables, such as the topological charge, which are notoriously affected by critical slowing down (CSD) in conventional algorithms \cite{schaeferCriticalSlowingError2011, allesHybridMonteCarlo1996}. A thorough investigation of these slow modes requires substantially larger statistics and lattice volumes, and consequently demands a fully parallelized implementation of the algorithm. Both the parallelization strategy and the subsequent analysis of non-local observables lie beyond the scope of this proof-of-concept study and will be addressed in a forthcoming publication.
	
	Finally, it is worth noting that ECMC and Heatbath exhibit comparable computational costs per sweep, as both algorithms involve numerically intensive operations -- including evaluations of trigonometric and exponential functions in addition to staple computations -- in contrast to the Metropolis algorithm, which relies solely on elementary arithmetic and staple calculations. As a result, reflective ECMC is less efficient than Metropolis in CPU time, while the forward variant stays better. Also, the efficiency advantage of the forward ECMC implementation over the reflective one is preserved in terms of CPU time, while the gap with Heatbath remains unchanged. As the present results were obtained using a basic implementation of ECMC, a thorough optimization of the ECMC sweep is expected to further improve its computational efficiency relative to classical algorithms in terms of CPU time.

	\section{Conclusion}\label{Conclusion}
	
	In this paper we have presented an alternative class of algorithms to simulate pure lattice gauge theories. So-called Event-Chain Monte-Carlo, they relax the detailed balance condition of the standard Markov-Chain Monte-Carlo algorithms like Metropolis or Heatbath. They respect the global balance and are rejection-free. The conceptual price to pay is an extension of the configuration space to a space with lift variables. They mix ballistic motion in configuration space with a stochastic evolution in the lift variables space. We were able to design two variants that are characterized by different transition rates in the lift variable space. They satisfy the invariance condition of PDMP and, thanks to symmetries of the Wilson gauge action, one of them is without backtracking. We have shown that distribution simulated by ECMC converge to the equilibrium one, whatever the bare coupling $\beta$ is. A numerical comparison of the plaquette autocorrelation time with the one obtained with Metropolis, Heatbath and Hybrid Monte-Carlo indicates that the forward variant of ECMC has a better efficiency than Metropolis and HMC in terms of CPU. In a companion paper we will address parallelization of ECMC and its ability to mitigate critical slowing down of the topological charge.
	
	\acknowledgments
	
	This work was performed using computational resources from the “Mésocentre” computing center of Université Paris-Saclay, CentraleSupélec and École Normale Supérieure Paris-Saclay\footnote{https://mesocentre.universite-paris-saclay.fr} supported by CNRS and Région Île-de-France. M.M. acknowledges the support of the French ANR under the grant ANR-20-CE46-0007 (\emph{SuSa} project).
	\newpage
	\appendix
	\section{Details of the runs}\label{sec:details}
	
	The numerical results for the mean plaquette $\langle P \rangle$ and the corresponding integrated autocorrelation times $\tau_{\mathrm{int}}$ are summarized in the following tables for different values of the coupling constant $\beta$. 
	
	For each algorithm, we report the estimated mean value where the statistical uncertainty is calculated using the $\Gamma$-method \cite{wolffMonteCarloErrors2004}. The efficiency of the sampling is quantified by $\tau_{\mathrm{int}}$, expressed in units of sweeps. 
	To verify the convergence and the statistical consistency of the various ECMC implementations (starting from both hot and cold configurations) against the reference Heatbath algorithm, we provide the $Z_H$-score. It is defined as:
	\begin{equation}
		Z_H = \frac{| \langle P \rangle_{\mathrm{algo}} - \langle P \rangle_{\mathrm{Heathbath}} |}{\sqrt{\sigma_{\mathrm{algo}}^2 + \sigma_{\mathrm{Heathbath}}^2}}
	\end{equation}
	where $\sigma$ denotes the statistical error of the respective algorithm. A $Z_H$-score significantly lower than 3 indicates that the results are statistically compatible, confirming that all implementations sample the correct distribution.
	
	\begin{table}[H]
		\centering
		\begin{tabular}{lllccc}
			\hline
			Algorithm & Start & Plaquette $\langle P \rangle$ & $\tau_{\mathrm{int}}$ & $Z_H$ & Samples \\
			\midrule
			Forward ECMC & Cold & 0.0601333(24) & 0.8621(67) & 1.09 & \num{768497} \\
			Forward ECMC & Hot & 0.0601300(11) & 0.8667(33) & 0.33 & \num{3548984} \\
			Refl. ECMC & Cold & 0.0601301(10) & 1.1713(42) & 0.37 & \num{5731472} \\
			Refl. ECMC & Hot & 0.0601293(14) & 1.1787(57) & 0.12 & \num{2967986} \\
			Heatbath & Cold & 0.0601288(32) & 0.5125(40) & - & \num{256500} \\
			\hline
		\end{tabular}
		\caption{Comparison for $\beta=1$.}
	\end{table}
	
	\begin{table}[H]
		\centering
		\begin{tabular}{lllccc}
			\hline
			Algorithm & Start & Plaquette $\langle P \rangle$ & $\tau_{\mathrm{int}}$ & $Z_H$ & Samples \\
			\midrule
			Forward ECMC & Cold & 0.1288111(27) & 1.385(12) & 0.15 & \num{1081496} \\
			Forward ECMC & Hot & 0.1288103(16) & 1.3819(69) & 0.04 & \num{3370985} \\
			Refl. ECMC & Cold & 0.1288118(15) & 2.0162(96) & 0.40 & \num{5583473} \\
			Refl. ECMC & Hot & 0.1288114(19) & 2.009(12) & 0.27 & \num{3218986} \\
			Heatbath & Cold & 0.1288105(30) & 0.5428(41) & - & \num{346500} \\
			\hline
		\end{tabular}
		\caption{Comparison for $\beta=2$.}
	\end{table}
	
	\begin{table}[H]
		\centering
		\begin{tabular}{lllccc}
			\hline
			Algorithm & Start & Plaquette $\langle P \rangle$ & $\tau_{\mathrm{int}}$ & $Z_H$ & Samples \\
			\midrule
			Forward ECMC & Cold & 0.2050398(44) & 2.663(33) & 0.26 & \num{909997} \\
			Forward ECMC & Hot & 0.2050468(23) & 2.694(18) & 1.41 & \num{3406484} \\
			Refl. ECMC & Cold & 0.2050442(23) & 4.156(28) & 0.75 & \num{5199974} \\
			Refl. ECMC & Hot & 0.2050419(31) & 4.129(37) & 0.16 & \num{2785487} \\
			Heatbath & Cold & 0.2050412(33) & 0.6342(54) & - & \num{378500} \\
			\hline
		\end{tabular}
		\caption{Comparison for $\beta=3$.}
	\end{table}
	
	\begin{table}[H]
		\centering
		\begin{tabular}{lllccc}
			\hline
			Algorithm & Start & Plaquette $\langle P \rangle$ & $\tau_{\mathrm{int}}$ & $Z_H$ & Samples \\
			\midrule
			Forward ECMC & Cold & 0.2904918(65) & 5.915(99) & 0.69 & \num{1050496} \\
			Forward ECMC & Hot & 0.2904971(36) & 5.778(56) & 0.01 & \num{3357485} \\
			Refl. ECMC & Cold & 0.2904986(36) & 9.536(92) & 0.29 & \num{5590473} \\
			Refl. ECMC & Hot & 0.2904953(49) & 9.45(12) & 0.29 & \num{2931487} \\
			Heatbath & Cold & 0.2904971(40) & 0.8422(88) & - & \num{391000} \\
			\hline
		\end{tabular}
		\caption{Comparison for $\beta=4$.}
	\end{table}
	
	\begin{table}[H]
		\centering
		\begin{tabular}{lllccc}
			\hline
			Algorithm & Start & Plaquette $\langle P \rangle$ & $\tau_{\mathrm{int}}$ & $Z_H$ & Samples \\
			\midrule
			Forward ECMC & Cold & 0.400444(10) & 19.58(41) & 2.09 & \num{2140990} \\
			Forward ECMC & Hot & 0.4004246(80) & 19.04(32) & 0.43 & \num{3412484} \\
			Refl. ECMC & Cold & 0.4004042(83) & 33.63(57) & 1.68 & \num{5648472} \\
			Refl. ECMC & Hot & 0.400421(12) & 34.67(81) & 0.04 & \num{2951486} \\
			Heatbath & Cold & 0.4004205(51) & 2.054(23) & - & \num{906997} \\
			\hline
		\end{tabular}
		\caption{Comparison for $\beta=5$.}
	\end{table}
	
	\begin{table}[H]
		\centering
		\begin{tabular}{lllccc}
			\hline
			Algorithm & Start & Plaquette $\langle P \rangle$ & $\tau_{\mathrm{int}}$ & $Z_H$ & Samples \\
			\midrule
			Forward ECMC & Cold & 0.594201(15) & 34.8(24) & 1.16 & \num{1376495} \\
			Forward ECMC & Hot & 0.5942349(94) & 34.0(16) & 1.32 & \num{3381485} \\
			Refl. ECMC & Cold & 0.594220(10) & 68.6(34) & 0.00 & \num{5692972} \\
			Refl. ECMC & Hot & 0.594214(17) & 94.4(73) & 0.30 & \num{2959486} \\
			Heatbath & Cold & 0.5942199(64) & 4.17(14) & - & \num{909496} \\
			\hline
		\end{tabular}
		\caption{Comparison for $\beta=6$.}
	\end{table}
	
	\begin{table}[H]
		\centering
		\begin{tabular}{lllccc}
			\hline
			Algorithm & Start & Plaquette $\langle P \rangle$ & $\tau_{\mathrm{int}}$ & $Z_H$ & Samples \\
			\midrule
			Forward ECMC & Cold & 0.6716472(71) & 9.50(35) & 0.10 & \num{908496} \\
			Forward ECMC & Hot & 0.6716447(35) & 8.88(18) & 0.35 & \num{3403984} \\
			Refl. ECMC & Cold & 0.6716410(39) & 18.11(40) & 1.02 & \num{5642972} \\
			Refl. ECMC & Hot & 0.6716382(54) & 18.09(53) & 1.26 & \num{2945487} \\
			Heatbath & Cold & 0.6716465(37) & 1.122(23) & - & \num{399998} \\
			\hline
		\end{tabular}
		\caption{Comparison for $\beta=7$.}
	\end{table}
	
	\begin{table}[H]
		\centering
		\begin{tabular}{lllccc}
			\hline
			Algorithm & Start & Plaquette $\langle P \rangle$ & $\tau_{\mathrm{int}}$ & $Z_H$ & Samples \\
			\midrule
			Forward ECMC & Cold & 0.7206891(48) & 7.34(13) & 0.52 & \num{1062496} \\
			Forward ECMC & Hot & 0.7206901(27) & 7.335(78) & 1.03 & \num{3441484} \\
			Refl. ECMC & Cold & 0.7206834(29) & 14.20(16) & 0.69 & \num{5625972} \\
			Refl. ECMC & Hot & 0.7206875(39) & 14.32(22) & 0.29 & \num{3185985} \\
			Heatbath & Cold & 0.7206862(28) & 0.9140(98) & - & \num{399998} \\
			\hline
		\end{tabular}
		\caption{Comparison for $\beta=8$.}
	\end{table}
	
	\begin{table}[H]
		\centering
		\begin{tabular}{lllccc}
			\hline
			Algorithm & Start & Plaquette $\langle P \rangle$ & $\tau_{\mathrm{int}}$ & $Z_H$ & Samples \\
			\midrule
			Forward ECMC & Cold & 0.7561657(46) & 6.71(14) & 0.04 & \num{776498} \\
			Forward ECMC & Hot & 0.7561646(22) & 6.731(69) & 0.40 & \num{3411484} \\
			Refl. ECMC & Cold & 0.7561677(24) & 12.71(14) & 0.54 & \num{5679972} \\
			Refl. ECMC & Hot & 0.7561697(33) & 12.54(18) & 0.95 & \num{2956486} \\
			Heatbath & Cold & 0.7561659(23) & 0.8443(87) & - & \num{399998} \\
			\hline
		\end{tabular}
		\caption{Comparison for $\beta=9$.}
	\end{table}
	
	
	\bibliographystyle{JHEP}
	\bibliography{biblio}
	
\end{document}